\documentclass[conference]{IEEEtran}
\IEEEoverridecommandlockouts
\usepackage{cite}
\usepackage{amsmath,amssymb,amsfonts}
\usepackage{algorithmic}
\usepackage{graphicx}
\usepackage{textcomp} 
\usepackage{xcolor}
\usepackage{url} 
\usepackage{xurl} 

\usepackage{tikz}
\usetikzlibrary{shapes.geometric, arrows.meta, positioning, fit, calc}
\usepackage{caption}

\def\BibTeX{{\rm B\kern-.05em{\sc i\kern-.025em b}\kern-.08em
    T\kern-.1667em\lower.7ex\hbox{E}\kern-.125emX}}

\tikzset{
    input/.style={rectangle, rounded corners, minimum width=2.4cm, minimum height=0.9cm, text centered, text width=2.2cm, draw=black, fill=blue!10, thick, font=\scriptsize},
    process/.style={rectangle, minimum width=2.8cm, minimum height=0.9cm, text centered, text width=2.6cm, draw=black, fill=green!10, thick, font=\scriptsize},
    llm_process/.style={rectangle, rounded corners, minimum width=2.8cm, minimum height=0.9cm, text centered, text width=2.6cm, draw=black, fill=orange!20, thick, font=\scriptsize},
    output_node/.style={rectangle, rounded corners, minimum width=2.4cm, minimum height=0.9cm, text centered, text width=2.2cm, draw=black, fill=red!10, thick, font=\scriptsize},
    data_item/.style={ellipse, minimum width=2.2cm, minimum height=0.8cm, text centered, text width=2.0cm, draw=black, fill=gray!5, font=\tiny, inner sep=1pt},
    arrow/.style={thick,->,>=Stealth},
    line_label/.style={font=\tiny, midway, sloped, fill=white, inner sep=0.5pt, above} 
}

\begin{document}

\title{LUST: A Multi-Modal Framework with Hierarchical LLM-based Scoring for Learned Thematic Significance Tracking in Multimedia Content}

\author{\IEEEauthorblockN{Anderson de Lima Luiz}
\IEEEauthorblockA{\textit{AImotion Bavaria} \\
\textit{Technische Hochschule Ingolstadt}\\
Ingolstadt, Germany \\
\texttt{anderson.delimaluiz@thi.de}}
}

\maketitle

\begin{abstract}
This paper introduces the Learned User Significance Tracker (LUST), a framework designed to analyze video content and quantify the thematic relevance of its segments in relation to a user-provided textual description of significance. LUST leverages a multi-modal analytical pipeline, integrating visual cues from video frames with textual information extracted via Automatic Speech Recognition (ASR) from the audio track. The core innovation lies in a hierarchical, two-stage relevance scoring mechanism employing Large Language Models (LLMs). An initial "direct relevance" score, $S_{d,i}$, assesses individual segments based on immediate visual and auditory content against the theme. This is followed by a "contextual relevance" score, $S_{c,i}$, that refines the assessment by incorporating the temporal progression of preceding thematic scores, allowing the model to understand evolving narratives. The LUST framework aims to provide a nuanced, temporally-aware measure of user-defined significance, outputting an annotated video with visualized relevance scores and comprehensive analytical logs.
\end{abstract}


\section{Introduction}
The exponential growth of video data \cite{ahluwalia2022comprehensive} necessitates sophisticated automated tools for content analysis and interpretation. A key challenge is the identification of segments that align with specific, often abstract or nuanced, user-defined themes or concepts of "significance". Traditional methods, often reliant on low-level feature matching or simple keyword spotting \cite{lew2006content}, may fall short in capturing the semantic depth and contextual dependencies inherent in such tasks.

To address this, this paper presents the Learned User Significance Tracker (LUST), a framework engineered to automatically identify and track user-defined thematic significance within video content over time. LUST's methodology is built upon the synergistic integration of multi-modal data processing \cite{baltrusaitis2018multimodal} and the advanced semantic reasoning capabilities of Large Language Models (LLMs) \cite{zhao2023survey, brown2020language}. The system's analysis is anchored by a user-provided textual reference summary, denoted $R_{sum}$, which articulates the specific theme or significance vector the user wishes to track.

The principal scientific contribution of LUST is its novel two-stage, LLM-driven relevance assessment architecture:
\begin{enumerate}
    \item \textbf{Direct Relevance Assessment:} For discrete video segments, an LLM evaluates the immediate relevance by jointly considering a representative visual frame $I_i$ (derived from $F_i$) and the contemporaneous transcribed speech $\mathcal{C}_{S,i}$ in relation to the user's defined theme $R_{sum}$, yielding a score $S_{d,i}$.
    \item \textbf{Contextual Relevance Assessment:} This stage refines the understanding of significance by prompting an LLM to consider the direct relevance score $S_{d,i}$ of the current segment in conjunction with a historical ledger of direct relevance scores $H'_{d,i-1}$ from preceding segments and the current segment's specific speech context $\mathcal{C}_{S,i}$. This allows LUST to model how significance evolves and is perceived within the broader temporal narrative of the video, resulting in a score $S_{c,i}$.
\end{enumerate}
This paper details the architectural components, the multi-modal processing pipeline, and the intricacies of the LLM-based scoring mechanisms that underpin the LUST framework.

\section{The LUST Framework: Detailed Methodology}

The LUST framework operates through a sequential pipeline, beginning with multi-modal input processing and culminating in LLM-driven relevance scoring and output generation. Each component is designed to extract and utilize information pertinent to assessing thematic significance as defined by the user. The primary inputs to the system are the video $V$ and the user's thematic reference summary $R_{sum}$. Key configurable parameters include the visual window duration $\Delta t_w$, speech context radius $\delta_t$, and the history length $N_{hist}$ for contextual scoring.

\begin{figure*}[htbp]
    \centering
    \begin{tikzpicture}[node distance=0.9cm and 1.2cm, scale=0.85, transform shape] 

        \node[input] (video_input) {Input Video ($V$)};
        \node[input, right=of video_input, xshift=8.5cm] (ref_summary_input) {User Reference Summary ($R_{sum}$)}; 

        \node[process, below=of video_input, yshift=-0.3cm] (segmentation) {Temporal Video Segmentation \& Repr. Frame Extraction};
        \node[data_item, below=of segmentation, yshift=-0.1cm] (rep_frames) {Repr. Frames ($I_i$)};

        \node[process, right=of segmentation, xshift=3.5cm] (audio_proc) {Audio Extraction ($A$) \& Standardization (WAV)}; 
        \node[process, below=of audio_proc, yshift=-0.1cm] (asr) {Automatic Speech Recognition ($M_{ASR}$)};
        \node[data_item, below=of asr, yshift=-0.1cm] (transcripts) {Timestamped Utterances ($U$)};

        \node[process, below=of rep_frames, xshift=3.5cm, yshift=-0.6cm, text width=3cm] (speech_aggregation) {Speech Context Aggregation for Visual Windows}; 
        \node[data_item, below=of speech_aggregation, yshift=-0.1cm] (speech_context) {Speech Context ($\mathcal{C}_{S,i}$)};

        \node[llm_process, below=of speech_context, yshift=-0.6cm, text width=3cm] (direct_llm) {Stage 1: Direct Relevance Scoring ($M_{LLM}$)};
        \node[data_item, below=of direct_llm, yshift=-0.1cm] (direct_scores) {Direct Relevance Scores ($S_{d,i}$)};

        \node[llm_process, below=of direct_scores, yshift=-0.6cm, text width=3cm] (contextual_llm) {Stage 2: Contextual Relevance Scoring ($M_{LLM}$)};
        \node[data_item, below=of contextual_llm, yshift=-0.1cm] (contextual_scores) {Contextual Relevance Scores ($S_{c,i}$)};

        \node[process, below=of contextual_scores, yshift=-0.6cm, text width=3cm] (output_gen) {Output Generation \& Visualization};
        
        \node[output_node, below=of output_gen, yshift=-0.3cm, xshift=-1.8cm] (annotated_video) {Annotated Video (with $S_{c,i}$ Overlay)};
        \node[output_node, right=of annotated_video, xshift=0.8cm] (log_files) {Analytical Log Files};

        \draw[arrow] (video_input) -- (segmentation);
        \draw[arrow] (video_input.east) -| (audio_proc.north);
        \draw[arrow] (segmentation) -- (rep_frames);
        \draw[arrow] (audio_proc) -- (asr);
        \draw[arrow] (asr) -- (transcripts);

        \draw[arrow] (rep_frames.south) -- ([yshift=-0.15cm]rep_frames.south) -| (speech_aggregation.west);
        \draw[arrow] (transcripts.south) -- ([yshift=-0.15cm]transcripts.south) -| (speech_aggregation.east);
        \draw[arrow] (speech_aggregation) -- (speech_context);

        \draw[arrow] (rep_frames.south) -- ([yshift=-0.6cm]rep_frames.south) -| (direct_llm.west) node[line_label, pos=0.25, anchor=east, yshift=-1.5mm] {Frame $I_i$};
        \draw[arrow] (speech_context.south) -- (direct_llm.north) node[line_label, pos=0.5, yshift=0.5mm] {Speech $\mathcal{C}_{S,i}$};
        \coordinate (ref_sum_direct_bend) at ($(ref_summary_input.south) + (0,-0.8cm)$);
        \draw[arrow] (ref_summary_input.south) -- (ref_sum_direct_bend) -| (direct_llm.east) node[line_label, pos=0.25, anchor=west, yshift=-1.5mm] {$R_{sum}$};
        \draw[arrow] (direct_llm) -- (direct_scores);

        \draw[arrow] (direct_scores.south) -- (contextual_llm.north) node[line_label, pos=0.5, yshift=0.5mm] {$S_{d,i}$ + $H'_{d,i-1}$};
        \coordinate (speech_context_fork) at ($(speech_context.south east)+(-0.4cm,-0.4cm)$);
        \draw (speech_context.south) -- (speech_context.south east) -- (speech_context_fork);
        \draw[arrow] (speech_context_fork) -| (contextual_llm.west) node[line_label, pos=0.8, anchor=east, yshift=-1.5mm] {Speech $\mathcal{C}_{S,i}$};
        
        \coordinate (ref_sum_context_bend) at ($(ref_summary_input.south) + (0,-2.8cm)$);
        \draw[arrow] (ref_summary_input.south) -- (ref_sum_context_bend) -| (contextual_llm.east) node[line_label, pos=0.25, anchor=west, yshift=-1.5mm] {$R_{sum}^{snip}$};
        \draw[arrow] (contextual_llm) -- (contextual_scores);
        
        \draw[arrow] (contextual_scores) -- (output_gen);
        \draw[arrow] (output_gen.south) -- ++(0,-0.2cm) -| (annotated_video.north);
        \draw[arrow] (output_gen.south) -- ++(0,-0.2cm) -| (log_files.north);

        \node[draw, densely dotted, inner sep=6pt, rounded corners, fit=(segmentation) (audio_proc) (asr) (rep_frames) (transcripts)] (preprocessing_box) {}; 
        \node[above left= 0.05cm of preprocessing_box.north east, font=\tiny\itshape, fill=white, inner sep=0.5pt] {Multi-modal Preprocessing}; 
        
        \node[draw, densely dotted, inner sep=6pt, rounded corners, fit=(direct_llm) (contextual_llm) (direct_scores) (contextual_scores) (speech_aggregation) (speech_context)] (scoring_box) {}; 
        \node[above left= 0.05cm of scoring_box.north east, font=\tiny\itshape, fill=white, inner sep=0.5pt] {Hierarchical LLM Scoring}; 

    \end{tikzpicture}
    \caption{Overall workflow of the LUST framework, illustrating the pipeline from multi-modal input processing ($V, R_{sum}$) to hierarchical LLM-based relevance scoring ($S_{d,i}, S_{c,i}$) and output generation.}
    \label{fig:lust_workflow}
\end{figure*}
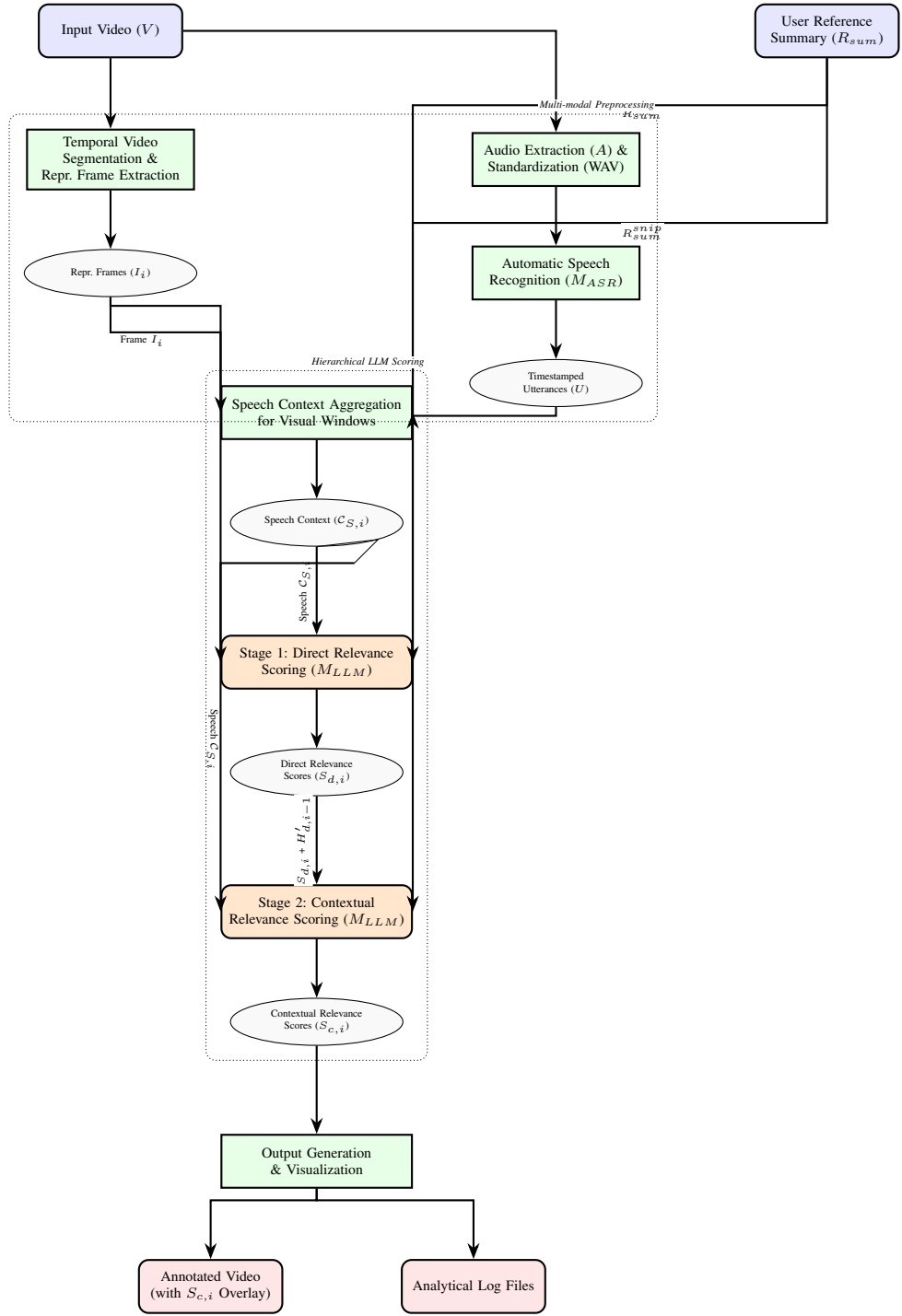

\subsection{Input Modalities and Preprocessing}
LUST processes two primary modalities from the input video $V$: visual and auditory. The user also provides a crucial textual input: the reference summary $R_{sum}$.

\subsubsection{User-Defined Thematic Anchor: The Reference Summary ($R_{sum}$)}
The reference summary, denoted $R_{sum}$, is a textual input provided by the user that describes the theme, concept, event, or pattern of interest they wish to track in the video. This summary serves as the semantic ground truth or query against which video segments are evaluated. Its quality and specificity directly influence the LLM's ability to identify relevant content. For example, $R_{sum}$ might be "tracking moments of escalating tension followed by a resolution" or "identifying instances of collaborative problem-solving".

\subsubsection{Temporal Video Segmentation and Visual Feature Extraction}
The input video $V$, with total duration $T_{vid}$, is first divided into $N_w = \lceil T_{vid} / \Delta t_w \rceil$ contiguous temporal segments, termed "visual windows", where $\Delta t_w$ is the configurable window duration (e.g., 1.0s).
For each visual window $i \in \{1, \dots, N_w\}$:
\begin{itemize}
    \item Start time: $t_{i}^{start} = (i-1) \cdot \Delta t_w$
    \item End time: $t_{i}^{end} = \min(i \cdot \Delta t_w, T_{vid})$
    \item Actual duration: $\Delta \tau_i = t_{i}^{end} - t_{i}^{start}$
    \item Center time: $t_{i}^{center} = t_{i}^{start} + \Delta \tau_i / 2$
\end{itemize}
The parameter $\Delta t_w$ dictates the granularity of the analysis. Within each visual window $i$, a single "representative frame" $F_i$ is selected, typically the frame temporally closest to $t_{i}^{center}$. This frame $F_i = \text{ExtractFrame}(V, t_{i}^{center})$ is converted into a PIL (Pillow) Image object \cite{pillow} and then encoded into a base64 data URI \cite{rfc2397}, denoted $I_i$, for transmission to the multi-modal LLM. The number of frames initially sampled within the window is a configurable parameter determining sampling density per second.

\subsubsection{Audio Processing and Automatic Speech Recognition (ASR)}
The audio track $A$ is demultiplexed from the video file $V$. A video processing utility (e.g., FFmpeg \cite{ffmpeg}) is used to convert $A$ into a standardized WAV format (16000 Hz, mono, 16-bit PCM signed little-endian). This standardized audio is then processed by an ASR system, $M_{ASR}$. LUST employs an efficient reimplementation of OpenAI's Whisper model \cite{openai_whisper_model_card_2022, radford2022robust}, with a configurable model size (e.g., a medium-sized English model). The ASR module, $M_{ASR}$, processes $A$ to produce a set of $N_u$ time-stamped utterances:
\begin{equation}
U = \{u_j = (t_{j}^{u,start}, t_{j}^{u,end}, \text{text}_j) \mid j=1, \dots, N_u\}
\label{eq:utterances}
\end{equation}
where $t_{j}^{u,start}$ and $t_{j}^{u,end}$ are the start and end times of utterance $j$, and $\text{text}_j$ is its transcribed content. Voice Activity Detection \cite{sohn1999statistical} is used to improve transcript quality.

\subsubsection{Speech Context Aggregation for Visual Windows}
To link spoken content with visual segments, LUST aggregates ASR-transcribed utterances relevant to each visual window $i$. The speech context $\mathcal{C}_{S,i}$ for window $i$ is formed by collecting utterances from $U$ that are temporally proximal to its center time $t_{i}^{center}$, within a configurable radius $\delta_t$ (e.g., 2.5s).
Let $\mathcal{J}_i = \{j \mid [t_{j}^{u,start}, t_{j}^{u,end}] \cap [t_{i}^{center} - \delta_t, t_{i}^{center} + \delta_t] \neq \emptyset \}$.
The speech context is then:
\begin{equation}
\mathcal{C}_{S,i} = \bigoplus_{j \in \mathcal{J}_i} \text{format}(u_j)
\label{eq:speech_context}
\end{equation}
where $\bigoplus$ denotes concatenation of formatted utterance strings (e.g., ``[\texttt{start\_time}s - \texttt{end\_time}s]: '\texttt{text}'``). If $\mathcal{J}_i = \emptyset$, $\mathcal{C}_{S,i}$ becomes a placeholder indicating no discernible speech. This $\mathcal{C}_{S,i}$ provides richer multi-modal context to the LLM.

\subsection{LLM-Powered Hierarchical Relevance Scoring}
The core intelligence of LUST resides in its two-stage relevance scoring process, utilizing a pre-trained Large Language Model $M_{LLM}$ (e.g., a model from the Mistral-Small series or similar \cite{mistralai2024mixtral}).

\subsubsection{LLM Configuration and Prompting Strategy}
All interactions with $M_{LLM}$ are governed by a global system prompt, $\Pi_{sys}$, which instructs the LLM on its role and desired output format (a numerical score [0.0, 1.0]). A low temperature parameter (e.g., 0.1) promotes deterministic outputs. The system employs distinct prompt templates for the different scoring stages and contexts. For direct relevance scoring, template $\mathcal{T}_{d,aud}$ is used when audio context is present, and template $\mathcal{T}_{d,vis}$ is used otherwise. For contextual relevance scoring of the initial segment, templates $\mathcal{T}_{c,init,aud}$ (with audio) and $\mathcal{T}_{c,init,vis}$ (without audio) are utilized. For subsequent segments, contextual scoring employs templates $\mathcal{T}_{c,hist,aud}$ (with audio) and $\mathcal{T}_{c,hist,vis}$ (without audio).

\subsubsection{Stage 1: Direct Relevance Scoring ($S_{d,i}$)}

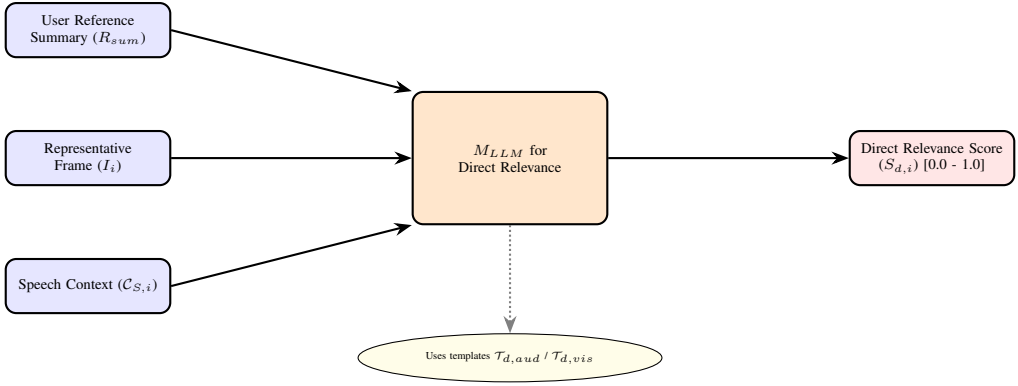
\begin{figure*}[htbp]
    \centering
    \begin{tikzpicture}[node distance=1.2cm and 1.5cm, scale=0.8, transform shape] 
        \node[input, text width=2.5cm] (ref_summary_s1) {User Reference Summary ($R_{sum}$)};
        \node[input, text width=2.5cm, below=of ref_summary_s1] (rep_frame_s1) {Representative Frame ($I_i$)};
        \node[input, text width=2.5cm, below=of rep_frame_s1] (speech_context_s1) {Speech Context ($\mathcal{C}_{S,i}$)};
        
        \node[llm_process, right=of rep_frame_s1, xshift=2.5cm, text width=3cm, minimum height=2.2cm] (llm_direct) {$M_{LLM}$ for Direct Relevance}; 
        
        \node[output_node, text width=2.5cm, right=of llm_direct, xshift=2.5cm] (direct_score_s1) {Direct Relevance Score ($S_{d,i}$) [0.0 - 1.0]}; 

        \draw[arrow] (ref_summary_s1.east) -- (llm_direct.north west);
        \draw[arrow] (rep_frame_s1.east) -- (llm_direct.west);
        \draw[arrow] (speech_context_s1.east) -- (llm_direct.south west);
        \draw[arrow] (llm_direct.east) -- (direct_score_s1.west);
        
        \node[data_item, text width=3.5cm, below=of llm_direct, yshift=-0.6cm, fill=yellow!10] (prompt_info_s1) {Uses templates $\mathcal{T}_{d,aud}$ / $\mathcal{T}_{d,vis}$};
         \draw[arrow, densely dotted, gray] (llm_direct.south) -- (prompt_info_s1.north);
    \end{tikzpicture}
    \caption{Stage 1: Direct Relevance Scoring. The LLM ($M_{LLM}$) assesses a video segment $i$'s relevance based on its representative frame $I_i$, associated speech context $\mathcal{C}_{S,i}$, and the user's reference summary $R_{sum}$, producing $S_{d,i}$. Prompt construction uses templates $\mathcal{T}_{d,aud}$ or $\mathcal{T}_{d,vis}$.}
    \label{fig:direct_relevance_scoring}
\end{figure*}

The first scoring stage assesses the immediate relevance of each visual window $i$. Let $P_{d,i}$ be the textual part of the user prompt for window $i$.
If speech context $\mathcal{C}_{S,i}$ is available (i.e., $\mathcal{C}_{S,i}$ does not indicate an absence of speech), $P_{d,i}$ is an instantiation of template $\mathcal{T}_{d,aud}$ using arguments $(R_{sum}, t_i^{start}, t_i^{end}, \mathcal{C}_{S,i})$.
Else, $P_{d,i}$ is an instantiation of template $\mathcal{T}_{d,vis}$ using arguments $(R_{sum}, t_i^{start}, t_i^{end})$.

The multi-modal input to $M_{LLM}$ consists of the image data URI $I_i$ and the textual prompt $P_{d,i}$. The direct relevance score $S_{d,i}$ is then given by:
\begin{align}
S_{d,i} = \text{clamp}_{[0,1]} (M_{LLM}(\text{user\_content} = [\{P_{d,i}\}, \{I_i\}], \nonumber \\ \text{system\_prompt}=\Pi_{sys}))
\label{eq:direct_relevance}
\end{align}
This score $S_{d,i}$ reflects the localized relevance of segment $i$ to $R_{sum}$.

\subsubsection{Stage 2: Contextual Relevance Scoring ($S_{c,i}$)}

\begin{figure*}[htbp]
    \centering
    \begin{tikzpicture}[node distance=1cm and 1.2cm, scale=0.8, transform shape] 
        \node[input, text width=2.2cm] (ref_summary_s2) {$R_{sum}^{snip}$};
        \node[input, text width=2.2cm, below=of ref_summary_s2] (past_scores_s2) {Score History ($H'_{d,i-1}$)};
        \node[input, text width=2.2cm, below=of past_scores_s2] (current_sd_s2) {Current $S_{d,i}$};
        \node[input, text width=2.2cm, below=of current_sd_s2] (speech_context_s2) {Current $\mathcal{C}_{S,i}$};
        
        \node[llm_process, right=of past_scores_s2, xshift=3cm, text width=3cm, minimum height=2.8cm] (llm_contextual) {$M_{LLM}$ (Text-only) for Contextual Relevance}; 
        
        \node[output_node, text width=2.2cm, right=of llm_contextual, xshift=3cm] (contextual_score_s2) {Contextual Relevance Score ($S_{c,i}$) [0.0 - 1.0]}; 

        \draw[arrow] (ref_summary_s2.east) -- (llm_contextual.north west);
        \draw[arrow] (past_scores_s2.east) -- (llm_contextual.west);
        \draw[arrow] (current_sd_s2.east) -- (llm_contextual.west);
        \draw[arrow] (speech_context_s2.east) -- (llm_contextual.south west);
        \draw[arrow] (llm_contextual.east) -- (contextual_score_s2.west);

        \node[data_item, text width=3.5cm, below=of llm_contextual, yshift=-0.6cm, fill=yellow!10] (prompt_info_s2) {Uses templates $\mathcal{T}_{c,init/hist,aud/vis}$};
        \draw[arrow, densely dotted, gray] (llm_contextual.south) -- (prompt_info_s2.north);
    \end{tikzpicture}
    \caption{Stage 2: Contextual Relevance Scoring. The LLM ($M_{LLM}$) refines relevance for segment $i$ by considering $R_{sum}^{snip}$, past scores $H'_{d,i-1}$, current $S_{d,i}$, and $\mathcal{C}_{S,i}$, yielding $S_{c,i}$. Prompt construction uses templates such as $\mathcal{T}_{c,init,aud}$, $\mathcal{T}_{c,hist,vis}$, etc.}
    \label{fig:contextual_relevance_scoring}
\end{figure*}
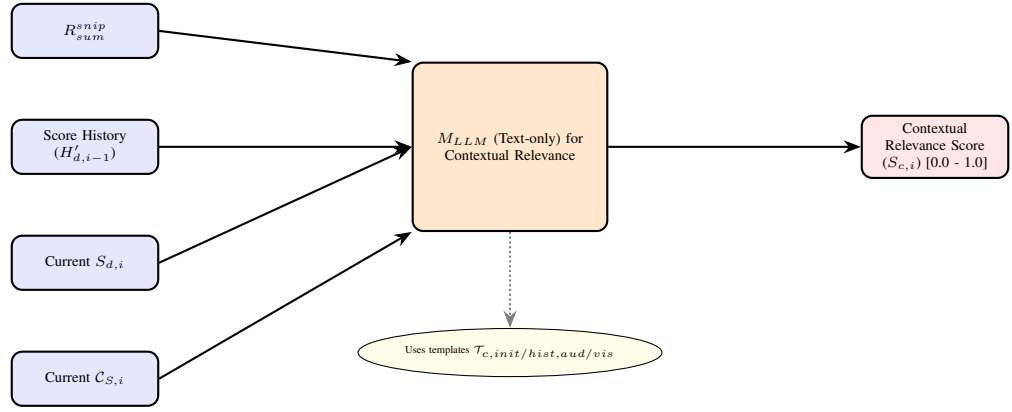

The second stage refines the assessment by incorporating temporal context, using only textual input for $M_{LLM}$. A snippet of the reference summary, $R_{sum}^{snip} = \text{truncate}(R_{sum}, L_{snip})$ (e.g., $L_{snip}=70$ characters), is used. The history of past direct scores up to window $i-1$ is $H_{d}^{(i-1)} = (S_{d,1}, \dots, S_{d,i-1})$. A truncated version for the prompt is $H'_{d,i-1} = (S_{d,k}, \dots, S_{d,i-1})$, where $k = \max(1, i - N_{hist})$ and $N_{hist}$ is the maximum number of past scores considered for the prompt. Let $H_{str,i-1}$ be the string representation of $H'_{d,i-1}$, and $I_{trunc}$ be an indicator if $H_{d}^{(i-1)}$ was truncated.

The textual user prompt for contextual relevance, $\Pi_{c,i}$, is constructed as follows:
\begin{itemize}
    \item For $i=1$ (initial window):
        If $\mathcal{C}_{S,i}$ is available: $\Pi_{c,i}$ is an instantiation of $\mathcal{T}_{c,init,aud}$ using arguments $(R_{sum}^{snip}, S_{d,i}, t_{i}^{start}, \mathcal{C}_{S,i})$.
        Else: $\Pi_{c,i}$ is an instantiation of $\mathcal{T}_{c,init,vis}$ using arguments $(R_{sum}^{snip}, S_{d,i}, t_{i}^{start})$.
    \item For $i > 1$:
        If $\mathcal{C}_{S,i}$ is available: $\Pi_{c,i}$ is an instantiation of $\mathcal{T}_{c,hist,aud}$ using arguments $(R_{sum}^{snip}, I_{trunc}, H_{str,i-1}, S_{d,i}, t_{i}^{start}, \mathcal{C}_{S,i})$.
        Else: $\Pi_{c,i}$ is an instantiation of $\mathcal{T}_{c,hist,vis}$ using arguments $(R_{sum}^{snip}, I_{trunc}, H_{str,i-1}, S_{d,i}, t_{i}^{start})$.
\end{itemize}
The contextual relevance score $S_{c,i}$ is then given by:
\begin{align}
S_{c,i} &= \text{clamp}_{[0,1]}\big(M_{LLM}(\text{user\_content} = [\Pi_{c,i}], \nonumber \\
&\hspace{7em} \text{system\_prompt} = \Pi_{sys})\big)
\label{eq:contextual_relevance}
\end{align}
This score $S_{c,i}$ represents a more nuanced understanding of segment $i$'s importance considering the evolving narrative.

\subsection{Output Generation and Visualization}
The LUST system generates several outputs.

\subsubsection{Data Logging and Archiving}
Comprehensive logs are saved for analysis and reproducibility:
\begin{itemize}
    \item A \textbf{configuration and summary log} is generated, recording the input video identifier, the reference summary $R_{sum}$, key processing parameters (such as $\Delta t_w$ and $N_{hist}$), and the overall average contextual score.
    \item The \textbf{full video transcription log} documents the complete set of timestamped utterances $U$ derived from the video's audio track.
    \item A detailed \textbf{segment analysis log} provides, for each visual window $i$, its temporal boundaries ($t_i^{start}, t_i^{end}$), the final contextual relevance score ($S_{c,i}$), the average direct relevance score ($S_{d,i}$), and the associated speech context snippet ($\mathcal{C}_{S,i}$).
    \item The representative visual frame $F_i$ (or its encoded version $I_i$) selected from each window is archived as an image file, often named to include its scores, facilitating qualitative review.
\end{itemize}
This detailed logging supports reproducibility, debugging, and deeper qualitative analysis of the system's performance.

\subsubsection{Video Overlay and Final Output}
\begin{figure}[h!]
    \centering
    \includegraphics[width=\linewidth]{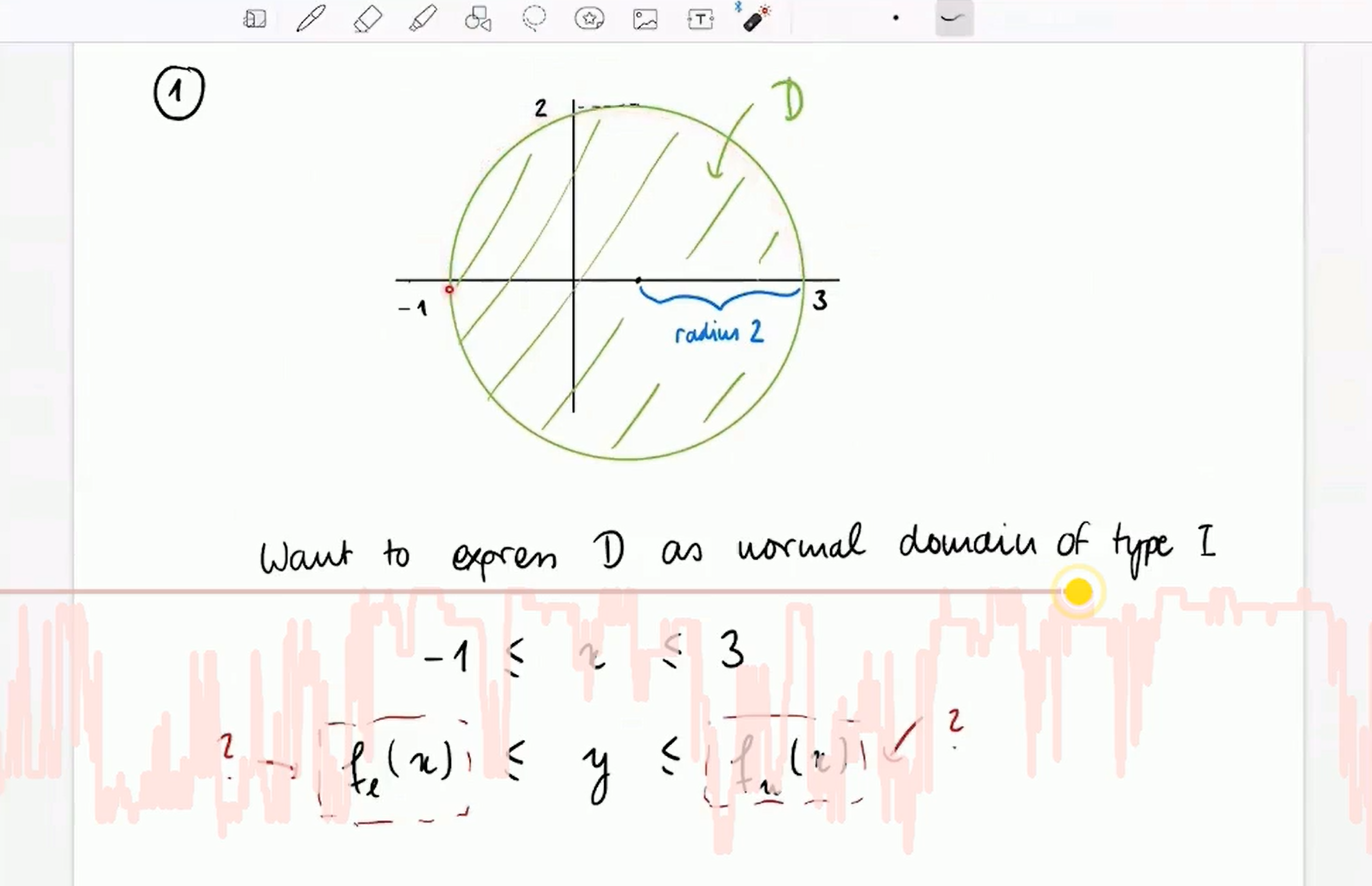}
    \caption{Demonstration of the thematic relevance (displayed in the overlay) for "Explanation of an example about calculation the integrals using a circle."}
    \label{LUST}
\end{figure}

A key output is an annotated version of the original video $V$, where the calculated contextual relevance scores ($S_{c,i}$) are visualized directly on the frames. This is achieved through:
\begin{enumerate}
    \item \textbf{Curve Generation:} The sequence of $S_{c,i}$ values for all visual windows $\{S_{c,i}\}_{i=1}^{N_w}$ is transformed into a set of 2D points. The x-coordinates correspond to temporal progression, and y-coordinates map $S_{c,i}$ values to a vertical range on the video frame. Cubic Bézier curves \cite{farin2002curves} are used for smooth interpolation between these points.
    \item \textbf{Overlay Drawing:} For each frame of the output video, the generated Bézier curve is drawn semi-transparently. A timeline and a prominent dot, indicating the current window's $S_{c,i}$ and moving along the curve, are also rendered.
\end{enumerate}
The original video frames, now with these overlays, are then re-encoded using a video processing utility (e.g., FFmpeg \cite{ffmpeg}), preserving the original audio track if available, into a final video file. This visual feedback mechanism allows users to intuitively identify and navigate to segments of high or low thematic relevance. Figure \ref{LUST} demonstrates the final view of an example math lecture.

\subsection{System Execution Considerations}
The system includes logic to select the optimal compute device (GPU or CPU) for ASR processing. The main function orchestrates the entire pipeline: system diagnostics, ASR model loading, audio extraction and transcription, iterating through visual windows for direct and contextual scoring, and finally, generating all output files including the annotated video.

\section{Discussion and Potential Applications}

The LUST framework, through its multi-modal analysis and hierarchical LLM-based scoring, offers a significant advancement in automated video content understanding. The distinction between direct ($S_{d,i}$) and contextual ($S_{c,i}$) relevance allows for a more sophisticated interpretation of thematic significance. By incorporating temporal context—how the relevance of previous segments $H'_{d,i-1}$ influences the perception of the current one—LUST can better model narrative structures and evolving themes.

The system's adaptability via the user-defined $R_{sum}$ makes it suitable for:
\begin{itemize}
    \item \textbf{Academic Research:} Analyzing ethnographic recordings for specific behaviors.
    \item \textbf{Media Production:} Locating B-roll or identifying narrative turning points.
    \item \textbf{Educational Content Analysis:} Pinpointing segments pertinent to learning objectives.
    \item \textbf{Content Moderation:} Aiding in identifying segments for review based on describable themes.
    \item \textbf{Market Research:} Analyzing focus group recordings for feature-specific discussions.
\end{itemize}

Performance is linked to $M_{ASR}$ and $M_{LLM}$ capabilities. Errors from $M_{ASR}$ can affect scoring. The LLM's interpretation of $R_{sum}$ and its scoring consistency are critical. The fixed duration $\Delta t_w$ and contextual history $N_{hist}$ may require tuning. Future research could explore adaptive windowing \cite{truong2007video}, more advanced temporal modeling (e.g., using recurrent neural networks \cite{donahue2015long} or transformers \cite{vaswani2017attention, bertasius2021space} over segment embeddings), incorporating explicit user feedback loops, and extending the framework to a wider range of LLMs. Investigating the interpretability of $M_{LLM}$ decisions \cite{danilevsky2020survey} also remains important.

\section{Conclusion}
The Learned User Significance Tracker (LUST) framework provides a robust and innovative approach to identifying and quantifying user-defined thematic relevance in video content. Its core strengths lie in its multi-modal data integration (visual $I_i$ and auditory $\mathcal{C}_{S,i}$) and its sophisticated two-stage LLM-based relevance scoring ($S_{d,i}$ and $S_{c,i}$). The hierarchical approach, from direct to contextual relevance, enables a nuanced interpretation of significance. The visualized $S_{c,i}$ scores on the output video make results accessible. LUST shows considerable potential for applications requiring deep semantic understanding of video narratives and user-specific thematic tracking.

\bibliographystyle{IEEEtran}
\bibliography{v3.bib} 

\begin{thebibliography}{10}
\providecommand{\url}[1]{#1}
\csname url@samestyle\endcsname
\providecommand{\newblock}{\relax}
\providecommand{\bibinfo}[2]{#2}
\providecommand{\BIBentrySTDinterwordspacing}{\spaceskip=0pt\relax}
\providecommand{\BIBentryALTinterwordstretchfactor}{4}
\providecommand{\BIBentryALTinterwordspacing}{\spaceskip=\fontdimen2\font plus
\BIBentryALTinterwordstretchfactor\fontdimen3\font minus \fontdimen4\font\relax}
\providecommand{\BIBforeignlanguage}[2]{{%
\expandafter\ifx\csname l@#1\endcsname\relax
\typeout{** WARNING: IEEEtran.bst: No hyphenation pattern has been}%
\typeout{** loaded for the language `#1'. Using the pattern for}%
\typeout{** the default language instead.}%
\else
\language=\csname l@#1\endcsname
\fi
#2}}
\providecommand{\BIBdecl}{\relax}
\BIBdecl

\bibitem{ahluwalia2022comprehensive}
P.~Ahluwalia and N.~Varshney, ``A comprehensive review on video summarization techniques,'' \emph{Artificial Intelligence Review}, vol.~55, no.~6, pp. 4455--4507, 2022.

\bibitem{lew2006content}
M.~S. Lew, N.~Sebe, C.~Djeraba, and R.~Jain, ``Content-based multimedia information retrieval: State of the art and challenges,'' \emph{ACM Transactions on Multimedia Computing, Communications, and Applications (TOMM)}, vol.~2, no.~1, pp. 1--19, 2006.

\bibitem{baltrusaitis2018multimodal}
T.~Baltru{\v{s}}aitis, C.~Ahuja, and L.-P. Morency, ``Multimodal machine learning: A survey and taxonomy,'' \emph{IEEE Transactions on Pattern Analysis and Machine Intelligence}, vol.~41, no.~2, pp. 423--443, 2019.

\bibitem{zhao2023survey}
W.~X. Zhao, K.~Zhou, J.~Li, T.~Tang, X.~Wang, Y.~Hou, Y.~Min, B.~Zhang, J.~Zhang, Z.~Dong, Y.~Du, C.~Yang, Y.~Chen, Z.~Chen, J.~Jiang, R.~Ren, Y.~Li, Z.~Liu, P.~Liu, J.-Y. Nie, and J.-R. Wen, ``A survey of large language models,'' \emph{arXiv preprint arXiv:2303.18223}, 2023.

\bibitem{brown2020language}
\BIBentryALTinterwordspacing
T.~B. Brown, B.~Mann, N.~Ryder, M.~Subbiah, J.~D. Kaplan, P.~Dhariwal, A.~Neelakantan, P.~Shyam, G.~Sastry, A.~Askell, S.~Agarwal, A.~Herbert-Voss, G.~Krueger, T.~Henighan, R.~Child, A.~Ramesh, D.~M. Ziegler, J.~Wu, C.~Winter, C.~Hesse, M.~Chen, E.~Sigler, M.~Litwin, S.~Gray, B.~Chess, J.~Clark, C.~Berner, S.~McCandlish, A.~Radford, I.~Sutskever, and D.~Amodei, ``Language models are few-shot learners,'' in \emph{Advances in Neural Information Processing Systems 33 (NeurIPS 2020)}, H.~Larochelle, M.~Ranzato, R.~Hadsell, M.~Balcan, and H.~Lin, Eds.\hskip 1em plus 0.5em minus 0.4em\relax Curran Associates, Inc., 2020, pp. 1877--1901. [Online]. Available: \url{https://proceedings.neurips.cc/paper/2020/file/1457c0d6bfcb4967418bfb8ac142f64a-Paper.pdf}
\BIBentrySTDinterwordspacing

\bibitem{pillow}
A.~Clark and Contributors, ``Pillow (pil fork),'' \url{https://python-pillow.org/}, 2024, accessed: 2024-06-04. Current version at access time: 10.3.0.

\bibitem{rfc2397}
\BIBentryALTinterwordspacing
L.~Masinter, ``The "data" url scheme,'' Request for Comments 2397, IETF, RFC 2397, Aug. 1998. [Online]. Available: \url{https://www.rfc-editor.org/info/rfc2397}
\BIBentrySTDinterwordspacing

\bibitem{ffmpeg}
{FFmpeg developers}, ``{FFmpeg Multimedia Framework},'' \url{https://ffmpeg.org}, 2024, accessed: 2024-06-04.

\bibitem{openai_whisper_model_card_2022}
{OpenAI}, ``Whisper model card,'' \url{https://github.com/openai/whisper}, 2022, accessed: 2024-06-04.

\bibitem{radford2022robust}
A.~Radford, J.~W. Kim, T.~Xu, G.~Brockman, C.~McLeavey, and I.~Sutskever, ``Robust speech recognition via large-scale weak supervision,'' \emph{arXiv preprint arXiv:2212.04356}, 2022.

\bibitem{sohn1999statistical}
J.~Sohn, N.~S. Kim, and W.~Sung, ``A statistical model-based voice activity detection,'' \emph{IEEE Signal Processing Letters}, vol.~6, no.~1, pp. 1--3, 1999.

\bibitem{mistralai2024mixtral}
{Mistral AI Team}, ``Mixtral of experts,'' \emph{arXiv preprint arXiv:2401.04088}, 2024.

\bibitem{farin2002curves}
G.~E. Farin, \emph{Curves and Surfaces for CAGD: A Practical Guide}, 5th~ed.\hskip 1em plus 0.5em minus 0.4em\relax San Francisco, CA, USA: Morgan Kaufmann Publishers Inc., 2002.

\bibitem{truong2007video}
B.~T. Truong and S.~Venkatesh, ``Video abstraction: A systematic review and classification,'' \emph{ACM Transactions on Multimedia Computing, Communications, and Applications (TOMM)}, vol.~3, no.~1, pp. 3--es, 2007.

\bibitem{donahue2015long}
J.~Donahue, L.~Anne~Hendricks, S.~Guadarrama, M.~Rohrbach, S.~Venugopalan, K.~Saenko, and T.~Darrell, ``Long-term recurrent convolutional networks for visual recognition and description,'' in \emph{Proceedings of the IEEE Conference on Computer Vision and Pattern Recognition (CVPR)}, 2015, pp. 2625--2634.

\bibitem{vaswani2017attention}
\BIBentryALTinterwordspacing
A.~Vaswani, N.~Shazeer, N.~Parmar, J.~Uszkoreit, L.~Jones, A.~N. Gomez, {\L}.~Kaiser, and I.~Polosukhin, ``Attention is all you need,'' in \emph{Advances in Neural Information Processing Systems 30 (NIPS 2017)}, I.~Guyon, U.~V. Luxburg, S.~Bengio, H.~Wallach, R.~Fergus, S.~Vishwanathan, and R.~Garnett, Eds.\hskip 1em plus 0.5em minus 0.4em\relax Curran Associates, Inc., 2017, pp. 5998--6008. [Online]. Available: \url{https://papers.nips.cc/paper/2017/file/3f5ee243547dee91fbd053c1c4a845aa-Paper.pdf}
\BIBentrySTDinterwordspacing

\bibitem{bertasius2021space}
\BIBentryALTinterwordspacing
G.~Bertasius, H.~Wang, and L.~Torresani, ``Is space-time attention all you need for video understanding?'' in \emph{Proceedings of the 38th International Conference on Machine Learning (ICML)}, ser. Proceedings of Machine Learning Research, M.~Meila and T.~Zhang, Eds., vol. 139.\hskip 1em plus 0.5em minus 0.4em\relax PMLR, 18--24 Jul 2021, pp. 813--824. [Online]. Available: \url{https://proceedings.mlr.press/v139/bertasius21a.html}
\BIBentrySTDinterwordspacing

\bibitem{danilevsky2020survey}
M.~Danilevsky, K.~Qian, R.~Aharonov, Y.~Katsis, B.~Kawas, and P.~Sen, ``A survey of the state of explainable ai for natural language processing,'' \emph{arXiv preprint arXiv:2010.00711}, 2020, presented at AACL-IJCNLP 2020.

\end{thebibliography}

\end{document}